\documentstyle[epsfig]{aipproc}
\begin{document}
\title{Status of VHE Astronomy c.2000}
\author{Trevor C. Weekes}
\address{Whipple Observatory, Harvard-Smithsonian Center for
Astrophysics,\\ P.O. Box 97, Amado, AZ 85645-0097 U.S.A.\\
e-mail: tweekes@cfa.harvard.edu}
\maketitle

\section{Retrospective}

As we enter a new century of high energy astrophysics, it behooves
us to consider briefly how far we have come and far we have yet to
go. It is some 47 years since Galbraith and Jelley mounted a ten
inch mirror in a garbage can with a photomultiplier at its focus
and under the impossibly damp skies of the Berkshires detected the
first Cherenkov light pulses from air showers \cite{G&J53}.
Within a decade the broad outlines of the atmospheric Cherenkov
technique, as it applied to the search for very high energy gamma
rays, had been defined and significant experiments built to pursue
these searches. Reading again these early papers \cite{G&J57}
\cite{chudakov65} \cite{J&P63} one cannot but marvel at the
prescience of the authors and the courage of the exponents who were
taking on what must have seemed an impossible task. It is amazing
how much was anticipated in these classic papers and how much is
"rediscovered" by later exponents of the technique with large teams
of experimenters, sophisticated telescopes, and vast simulations.
Using simple, but elegant, telescopes, analytical models, and
considerable foresight, much of what was later to become the basis
of the multi-million dollar observatories now in operation was
anticipated. The early pioneers of the field, W. Galbraith and
J.V. Jelley of the Atomic Energy Research Establishment, Harwell,
England, N.A. Porter of University College, Dublin, Ireland, A.E.
Chudakov and V.I. Zatsepin of the Lebedev Institute, Moscow,
U.S.S.R.,  were the giants on whose work all future progress
depended; their names should be inscribed large within the High
Energy Astrophysics Hall of Fame! It is but an accident of nature
that these early experiments failed to detect significant signals
from supernova remnants and quasars since their sensitivities were
remarkably close to subsequent detection levels. 

To consider the state of high energy astrophysics in
1960 is to appreciate their courage in undertaking very high energy
gamma-ray experiments at that time. In fact "high energy
astrophysics" per se did not exist as a discipline at that
time. Although the seminal papers containing estimates of fluxes of
100 MeV and TeV gamma-rays to be expected from cosmic sources were
already in print \cite{morrison58} \cite{cocconi60}, they were,
with hindsight, hopelessly
optimistic. One can only hope that contemporary estimates of TeV
fluxes of neutrinos from cosmic sources are on a more solid base!
Gamma-ray astronomy at any energy was in its infancy with no
sources detected. X-ray astronomy did not exist and was
not even seriously considered. Since cosmic rays were both the
prime motivator to look for gamma-ray sources and the chief source
of information as to what properties such sources might have, the
conventional wisdom was that the dominant energy spectrum would be
that of the cosmic radiation, i.e. a differential exponent of -2.7.
Such an exponent did not bode well for the chances of detecting
very high energy gamma-rays. Since space gamma-ray telescopes could
be shielded with active detectors to exclude the charged cosmic
radiation, the chances of detecting a signal in a non-shielded
ground-based detector seemed vanishingly small. Optical detection
methods which depended on an unstable atmosphere and on the
detection of a weak optical signal against a background of natural
and man-made light sources did not seem promising. Only a fool or
someone with great courage would choose to pursue such techniques
in the face of such uncertainty.

The eventual detection of a mixed population of very high energy
(VHE) gamma-ray sources is a vindication of these early efforts and
a tribute to the pioneers.

\section{Gamma-ray Astronomy's Great Failure!}

Although GeV-TeV gamma-ray astronomy has had a number of
outstanding successes (the detection of blazars, the GEV component
in solar flares, the GeV-TeV component of GRBs, pulsars, shell and
plerionic supernova remnants, the galactic plane, to mention but a
few), the single great motivator, the conclusive solution to the
problem of the origin of the cosmic radiation, is still elusive. It
was this problem more than any other that led to the development of
high energy gamma-ray astronomy, both in space and on the ground,
but in some ways, at least observationaly, we are no closer to
identifying the source than we were 40 years ago. While we can
celebrate the contributions of gamma-ray studies to pulsar
phenomenology, to limiting the infrared background, to the study of
jets in extragalactic sources, we cannot yet proclaim with any
confidence  "Gamma-ray astronomers find the Origin of the Cosmic
Radiation".

As pointed out previously \cite{weekes99} every gamma-ray source
detected to date can be explained as a source of cosmic electron
acceleration and interaction; there is no source that can be
conclusively attributed to hadron acceleration. The $\pi$$^o$ bump
is only observed in the galactic plane and there we observe the
propagation, not the source, of the cosmic rays.
 
The title to this section is deliberately provocative and hopefully
will be obsolete before this volume is off the press. The number of
sensitive ground-based telescopes now on-line, or shortly
to come on-line, ensures that a vast number of new candidate
sources
are under ever increasingly sensitive scrutiny and may soon supply
the vital data necessary to solve this problem.

\section{What the Imaging A.C.T. has done for Astrophysics:}

\subsection{Existence of Sources}

\begin{figure}
\centerline{\epsfig{file=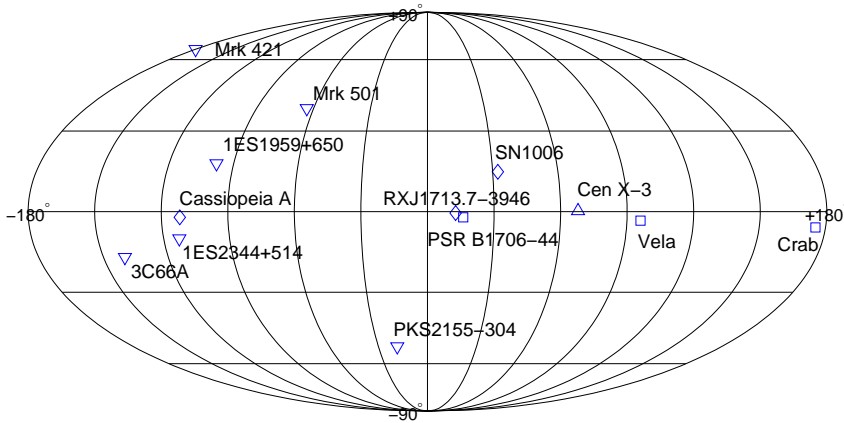,width=4.5in,angle=0.}}
\caption{Map of the sky at TeV energies in galactic coordinates
showing 13 discrete sources  (from
R.W.Lessard, private communication)
}
\label{skymap}
\end{figure}

The outstanding contribution of the imaging atmospheric Cherenkov
technique (IACT) must surely be that it has elevated ground-based
gamma-ray astronomy techniques above that critical threshold where
sources are detected with some credibility. In short the IACT has
confirmed that there is a gamma-ray sky at energies above 200 GeV
(Figure~\ref{skymap}), that the upper energy limit of space 
gamma-ray telescopes is an instrumental limit, not an astrophysical
one,
that there are a variety of different kinds of source populations,
that there are both galactic and extragalactic sources, both steady
and variable sources, and both point-like and extended. Furthermore
the energy spectra extend to 50 GeV where the flux sensitivity of
the IACT is not large so that the construction of new telescopes
that are sensitive above this energy is now justified. 

Half of the reported sources are galactic. These include plerions,
shell SNRs in which the progenitor particles are probably electrons
and an X-ray binary.
The extragalactic sources are all blazars. All but one are 
X-ray-selected BL Lacs; this population 
is quite different from the population
of EGRET-detected blazars.

Although a large fraction of the sky has not been surveyed with
high sensitivity, there are a number of objects for which
significant upper limits have been established. These have the
effect of severely limiting some classes of models and thus
contribute to high energy astrophysics in a negative sense. These
non-sources include shell-type SNRs, in which the progenitor
particles are hadrons, the galactic plane and various pulsars.

The energy range covered by atmospheric Cherenkov telescopes
extends from 50 GeV to 50 TeV. Thus far, the IACT has only been
used
down to 200 GeV but it is possible to reach as low as 50 GeV using 
Solar Arrays. By observing at low elevations the IACT can reach
energies as high as 50 TeV. 

\vspace*{-0.2cm}
\subsection{Source Catalog}

It is a matter of some disappointment (given the number of new
observatories now in operation) that the source catalog
(Table \ref{catalog}) shows no changes since the one assembled at
the
time of the Snowbird Workshop \cite{weekes99}. In particular the
credibility rating of sources has not changed.

\begin{center}
\begin{table}
\caption{Source Catalog c.2000 (Heidelberg)}
\label{catalog}
\begin{tabular}{llllll}

Source &   Type & z &   Discovery & EGRET & Grade  \\
\hline
{\bf Galactic Sources} &    &           &              &         & 
        \\

Crab Nebula &       Plerion &  &     1989 &      yes &   A  \\

PSR 1706-44 &       Plerion? &  &   1995 &  no &    A
\\

Vela &              Plerion? & &    1997 &  no &    B
\\

SN1006 &            Shell &  &      1997 &  no &   B$-$ \\

RXJ1713.7-3946 &    Shell &   &     1999 &  no &   B \\

Cassiopeia A   &    Shell &  &    1999 &  no &   C \\

Centaurus X-3  &    Binary &  &      1999 &  yes &  C \\

{\bf Extragalactic Sources} &  &         &              &    &    
 
\\

Markarian 421  &    XBL & 0.031 &  1992 &  yes &  A \\

Markarian 501  &    XBL & 0.034 & 1995 &  yes &  A \\

1ES2344+514    &    XBL & 0.044 & 1997 & no  &   C \\

PKS2155-304    &    XBL & 0.116 & 1999 & yes &  B \\

1ES1959+650    &    XBL & 0.048 & 1999 &  no &   B$-$ \\

3C66A          &    RBL & 0.44 &  1998 &  yes & C \\

\end{tabular}
\end{table}
\end{center}

\vspace*{-0.2cm}
\subsection{Time Variability}
One of the extraordinary contributions of VHE gamma-ray
observations to high energy astrophysics has been the revelation of
the variability of blazars on a wide range of time-scales. This has
truly opened a new window to the study of these objects and has
been a major incentive to improve and extend the observations.

\begin{figure}
\centerline{\epsfig{file=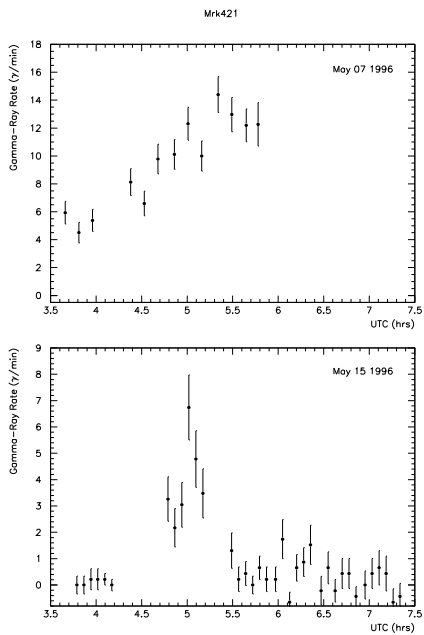,width=2.5in,angle=0.}}
\caption{Examples of short flares from Markarian 421 in May, 1996
reported by the Whipple Observatory \protect\cite{gaidos96}.}
\label{shortflare}
\end{figure}

{\bf Short-term Variations:} The dramatic flare  from Markarian 421
observed by the Whipple group on May 15, 1996 \cite{gaidos96} is
still the best instance of a rapid time variation (doubling time
$<$
15 minutes) from an AGN (Figure~\ref{shortflare}). The observation
of such short flares from more distant AGN may have some important
consequences for cosmology and quantum gravity \cite{biller99}.


\begin{figure}
\centerline{\epsfig{file=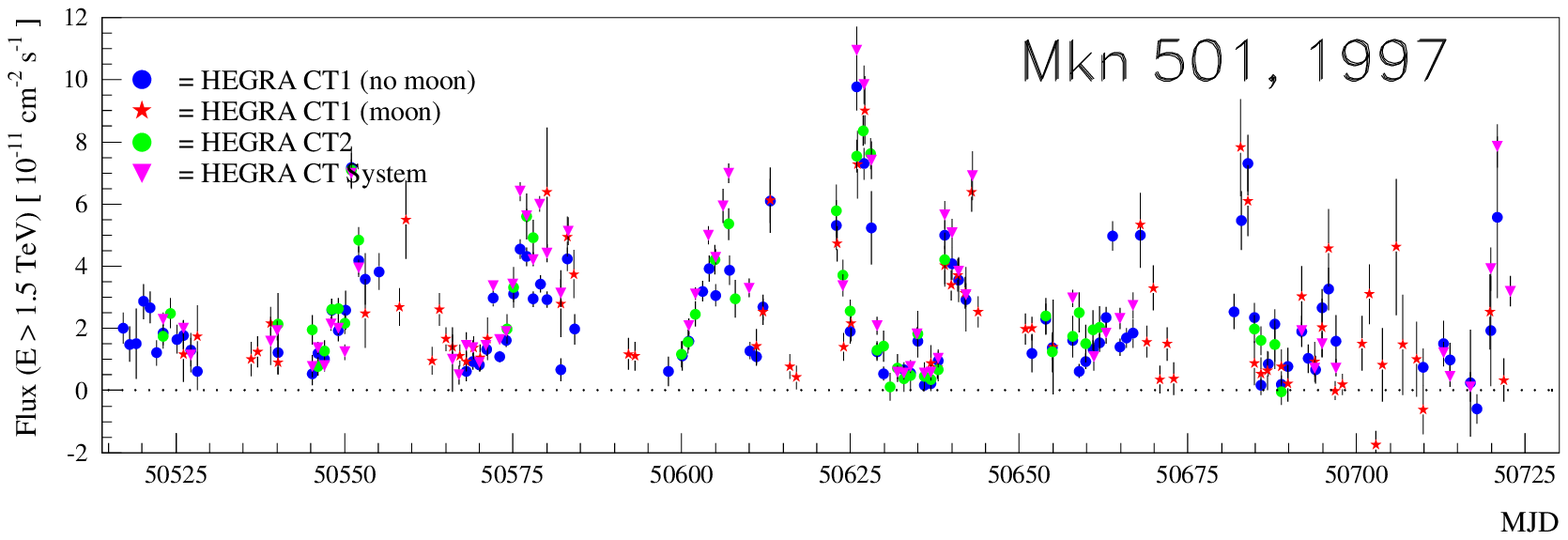,width=5in,angle=0.}}
\vspace*{0.2cm}
\caption{The light curve from Markarian 501 as seen at TeV energies
with the HEGRA telescopes \protect\cite{kranich99}.}
\label{hegra1997}
\end{figure}

{\bf Longterm Monitoring:} The ability of VHE telescopes to provide
longterm monitoring of variable sources is a little appreciated
property of ground-based observatories. Although the optical
techniques have a limited duty-cycle, their ability to monitor
sources over days, weeks, months and years is a unique feature in
gamma-ray astronomy above 1 MeV. This is dramatically illustrated
by the HEGRA observations of Markarian 501 in 1997
(Figure~\ref{hegra1997}). In this instance the observations were
significantly extended by the pioneering work of the HEGRA group in
observing in moonlight \cite{kranich99}. Such long-term
monitoring can be enhanced by organized observing campaigns
involving VHE observatories at different longitudes
\cite{takahashi00}; unfortunately the number of really sensitive
VHE telescopes is limited so that it is difficult to get continuous
coverage. Such campaigns are further limited by weather patterns.
Near continuous coverage will come with the next generation of
space telescopes which will use wide field telescopes.


\subsection{Multiwavelength Observations}

\begin{figure}
\centerline{\epsfig{file=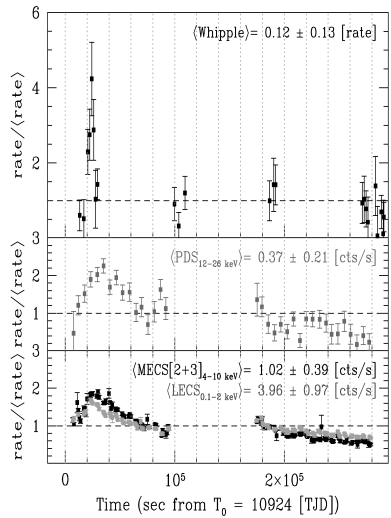,width=2in,angle=0.}}
\caption{The light-curve of TeV gamma-rays from Markarian 421 seen
at the Whipple Observatory during a multi-wavelength campaign with
the BeppoSAX X-ray satellite in April, 1998
\protect\cite{maraschi99}}
\label{maraschi-fig}
\end{figure}

For an understanding of the mechanisms at work in AGN jets,
multiwavelength campaigns are required. These are notoriously
difficult to organize since many traditional observatories require
advanced notification and approval and do not readily respond to
Targets of Opportunity. VHE observatories have the advantage that
they are usually controlled by the principal investigators, do not
have rigid observing schedules, and have only a small number of
known sources (regrettably!). The correlation of X-ray and VHE
observations has been reported in a number of instances
(cf.~\cite{cataneseweekes99} and references therein). One such
instance is shown in Figure~\ref{maraschi-fig}.

X-ray observations of sources with non-thermal spectra have proved
particularly useful in identifying candidate VHE supernova
remnants.
Hard X-ray observations hold the promise of identifying a new class
of extreme blazars which will have detectable and variable TeV
emission. The launch of EXIST, the first hard X-ray survey
instrument, in the next decade will extend this symbiotic
relationship.


\subsection{Energy Spectrum}

\begin{figure}
\centerline{\epsfig{file=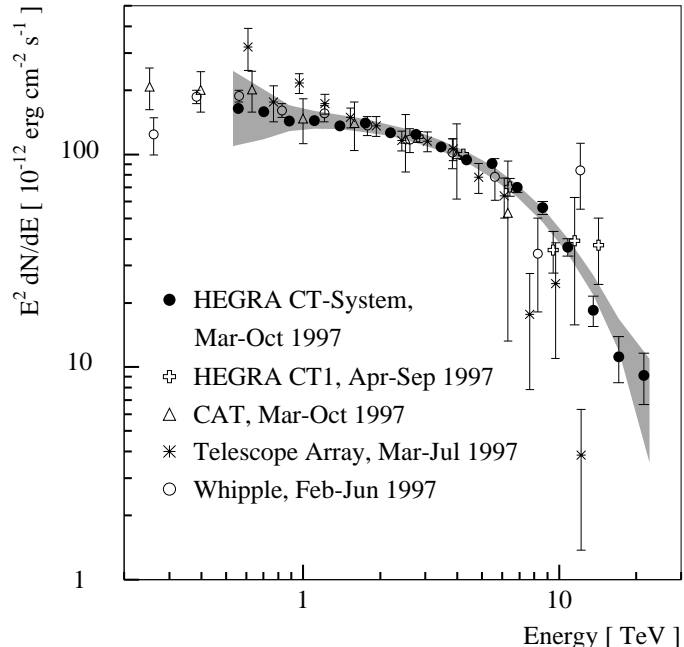,width=3.5in,angle=0.}}
\caption{Energy spectrum of Markarian 501 as reported by HEGRA,
CAT, 
7TA and Whipple \protect\cite{aharonian99}}
\label{aharonian-fig}
\end{figure}

The ability of the IACT to measure energy spectrum in the TeV
region has steadily improved and is comparable or better than that
achieved in space gamma-ray telescopes. Energy resolutions of
single telescope systems are quoted as 35\% and for the HEGRA
array as low as 10-15\%. It is reassuring that observations made
by different experiments using different methods of analysis are in
good agreement (Figure~\ref{aharonian-fig}).

\section{Status of VHE Observatories}

The IACT continues to be the favored method of detecting gamma-rays
in the 100-10,000 GeV energy range although there are several
overlapping techniques at the low and high energy ends of this
range. The principal observatories who have reported results using
this technique at recent workshops are listed (Table \ref{IACT}). 
Sadly at this meeting we learnt that both the Durham and 7TA
observatories, after the successful detection of  several sources,
have ceased operations. Also listed (Table \ref{NIACT}) are several
atmospheric Cherenkov observatories which do not use imaging; the
Potchefstroom experiment has also recently ceased operation.
 
In the interval between the death of EGRET and the launch of GLAST
it is interesting to explore the lowest energies that can be
achieved from the ground. Although it is unlikely that ground-based
observations can be really competitive below 50 GeV, there is a
window of opportunity between 10 and 100 GeV before the next
generation of space telescopes come into operation. Even relatively
simple telescopes could elucidate such problems as pulsar spectral
cut-offs, AGN absorption in intergalactic space, gamma-ray bursts,
etc. The simplest approach is to use large mirror collection areas
and these are provided by the relatively crude optics of large
solar collectors using heliostats. The four experiments listed in
Table \ref{SOLAR} use existing solar farms (large fields of
heliostats pointing, and roughly focussing, light to a central
tower). STACEE uses a facility that is still in operation for
experimental solar energy work whereas the other three use
facilities that are no longer in use. 

The energy threshold of particle air shower detectors 
(Table \ref{EAS}) has gradually
been reduced to the point where there is now overlap with IACT
experiments. The Tibet experiment has been in operation for some
years whereas MILAGRO has just come on-line.
\begin{center}
\begin{table}
\caption{Atmospheric Cherenkov Imaging Observatories c. June
2000}
\label{IACT}
\begin{tabular}{llcccl}

Group/ &   Location & Telescope(s) & Camera & Threshold 
& Ref. \\
Countries  & & & & & \\
 &     &  & Num.$\times$Apert.&  Pixels & (TeV)  \\ \\
\hline
Whipple & Arizona & 10\,m & 490 & 0.25 &
\cite{WHIPPLE} \\
USA-Ireland-UK & & & & & \\
Crimea  & Crimea & 6$\times$2.4\,m & 6$\times$37 & 1 
& \cite{CRIMEA} \\
Ukraine & & & & &  \\
SHALON & Tien Shen & 4\,m & 244 & 1.0 &
\cite{SHALON} \\
Russia & & & & &  \\
CANG.-II & Woomera & 10\,m & 256 & 0.5 
& \cite{CANGAROO}\\
Japan-Aust. & & &  & & \\
HEGRA &  La Palma. & 6$\times$3\,m &
5$\times$271  & 0.5 & \cite{HEGRA} \\
German.-Sp.Arm. &  & & & & \\
CAT &  Pyren\'ees & 4.5\,m & 600 & 0.25 & \cite{CAT}  \\
France & & & & &  \\
TACTIC & Mt.Abu & 10\,m & 349 & 0.3 & \cite{TACTIC}
\\
India & & & & &  \\
Durham & Narrabri & 3$\times$7\,m & 1$\times$109 & 0.25
& \cite{DURHAM99} \\
Uk & & & & &  \\
7TA &  Utah & 7$\times$2\,m & 7$\times$256 & 0.5 
& \cite{7TEL99} \\
Japan & & & & & \\
\end{tabular}
\end{table}
\end{center}

\begin{center}
\begin{table}
\caption{Non-imaging VHE Observatories c. June 2000}
\label{NIACT}
\begin{tabular}{lllcl}

Group & Countries & Type & Telescopes & Ref. \\
\hline 
Potchefstroom & South Africa & Lateral Array  & 4 & \cite{POTCH} \\
\\
Pachmarhi & India & Lateral Array & 25 & \cite{PACHMARI99} \\ \\
Beijing & China & Double & 2 & \cite{BEIJING} \\ \\
\end{tabular}
\end{table}
\end{center}

\begin{center}
\begin{table}
\caption{Atmospheric Cherenkov Solar Array Telescopes c. June
2000} 

\label{SOLAR}
\begin{tabular}{cccccl}

Group & Countries & Location & Heliostats & Threshold & Ref.
\\
     &          &   &    Now (future)  & GeV & \\
\hline                   
STACEE & Canada-USA & Albuquerque, USA &  32 (48)  & 180
& \cite{STACEE99} \\ \\
CELESTE & France & Themis, France & 40 (54) &     50$\pm10$ &
\cite{CELESTE99} \\ \\
Solar-2 & USA & Barstow, USA & 32 (64) &   20 & \cite{SOLAR-299} \\
\\
GRAAL & Germany-Spain & Almeria, Spain & (13-18)x4 & 200 GeV &
\cite{GRAAL99} \\ \\
\end{tabular}
\end{table}
\end{center}

\begin{center}
\begin{table}
\caption{Non-Atmospheric Cherenkov VHE Telescopes c. June
2000}
\label{EAS}
\begin{tabular}{lllllll}

Group & Countries & Location & Telescope & Altitude & Threshold 
& Ref. \\
     &          &   &           &      km      & TeV &  \\
\hline                   
Milagro & USA & Fenton Hill, NM & Water Cher. & 2.6 & 0.5 &
\cite{MILAGRO} \\ \\
Tibet HD & China-Japan & Tibet & Scintillators & 4.5 & 3 &
\cite{TIBET99}  \\ \\
\end{tabular}
\end{table}
\end{center}

\section{Next Generation Telescopes}

The next few years will see the completion of several new "next
generation" IACTs which will significantly improve the scientific
potential of the discipline. These major observatories will
probably dominate the field for the next decade and they represent
a major transition from the traditional "small science" which
characterized the early years of the IACT to multi-national
facilities which will serve a larger community as guest
investigators. Three of the facilities: CANGAROO-III
\cite{mori99}, HESS \cite{HESS99}, and VERITAS
\cite{VERITAS99}, build on the IACT array concept that has
been demonstrated by HEGRA and are very similar in concept; the
first two will be in the southern hemisphere, the third in the
northern hemisphere. MAGIC is a single large imaging telescope
which will use several new technological approaches \cite{MAGIC99}.
Some of the most important parameters of these telescopes are
listed in Table~\ref{telescopes}. A new addition to the list of new
IACTs is MACE, an Indian look-alike of MAGIC; its parameters were
described at this meeting.

\begin{center}
\begin{table}
\caption{MAGIC, HESS, CANGAROO-III and VERITAS}
\label{telescopes}
\begin{tabular}{lllll} 
{\bf Parameter}& MAGIC & HESS & CANGAROO-III & VERITAS \\ 
\hline
{\bf Base } & Munich & Heidelberg & Tokyo &
Arizona \\
{\bf Country} & Germany & Germany & Japan & U.S.A. \\ 
{\bf Partners} & Spain, Italy & France & 
Australia & UK, Ireland \\ 
{\bf Science} & AGNs, Bursts & SNR & Gal. Sources & AGNs,
SNR, Bursts \\ 
{\bf Location} & La Palma & Namibia & Woomera & Arizona
\\ 
{\bf Elevation}& 2.3 km & 1.8 km & S.L. & 1.4 km \\ 
{\bf \# of tel.}& 1 & 4 (16) & 4 & 7 \\ 
{\bf Pattern} & - & Square & Square & Hexagon \\
{\bf Spacing}& - & 120m & 100m & 80m  \\
{\bf Design}& Parabola & Davies-Cotton & Parabola & Davies-Cotton
\\ 
{\bf Aperture}& 17m & 12m & 10m & 10m\\ 
{\bf Focal length}& 20m & 15m & 8m & 12m \\
{\bf OSS} & Carbon fiber & Steel & Steel & Steel  \\
{\bf Facets }& 60cm square & 60cm circ. & 80cm circ. & 60cm
hex. \\ 
{\bf Material} & Al-milled & Ground-glass & Composite &
Glass
\\ 
{\bf Supplier} & Italy & Czech/Arm. & Japan & USA  \\
{\bf PMTs}& EMI & Phillips & Hama. & Hama.?  \\
{\bf Cabling}& Fiber & Coax & Coax & Coax \\ 
{\bf Electronics}& FADC & - & - & FADC  \\
{\bf \# of pixels }& $>$800 & 800 x 4 & 512 x 4 & 499 x 7   \\
{\bf First light} & 2001 & 2002 & 2003 & 2005 \\ 

\end{tabular}
\end{table}
\end{center}

{\bf Sensitivity:}
New experiments are often characterized by their energy thresholds
and flux sensitivities. The definitions of these quantities are not
trivial and often cause confusion. Energy threshold, usually
defined as the maximum in the differential response curve for a
Crab-like spectrum, is particularly misleading since the event
selection can be biassed to give a very low energy threshold but
with very small collection area.

\begin{figure}
\centerline{\epsfig{file=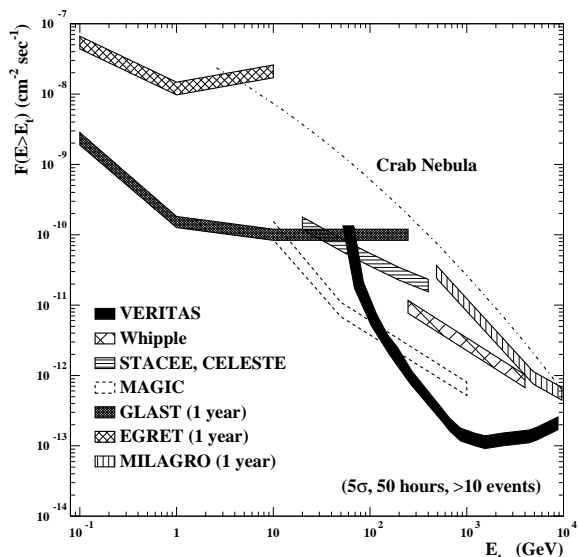,width=3in,angle=0.}}
\caption{Comparison of the point source sensitivity of VERITAS to
Whipple, MAGIC, CELESTE/STACEE/Solar-2,  EGRET, and MILAGRO. 
 }
\label{sensitivity}
\end{figure}

Figure~\ref{sensitivity} gives the integral flux sensitivity for
various existing experiments as well as the predicted sensitivity
for GLAST and VERITAS \cite{vassiliev99}. This figure has been
widely reproduced. However integral flux sensitivities presuppose
knowledge of the source spectrum and are not useful for sources
with steep spectra which fall off near the energy threshold. The
flux sensitivity for VERITAS is in close agreement with that
derived for HESS \cite{hofmann99} as would be expected. However
the sensitivity quoted for CANGAROO-III \cite{mori99} is much
better, particularly at lower energies; this prediction needs to
be reconciled with those of VERITAS and HESS since the array
parameters are not that different.


{\bf Schedule:}
Of major concern to all those interested in VHE astronomy is the
roadmap of HE and VHE experiments in the next decade. Although all
proposed launch and construction completion dates are inevitably
optimistic, it is hoped that the future depicted in \cite{weekes99}
is not grossly inaccurate. The solar telescopes are
already reporting results. The MAGIC web page proudly, if somewhat
optimistically, gives a countdown to first light in 2001. Concrete
has already been poured for the HESS foundations and one CANGAROO
telescope is in place and taking data. Although the first of the
next generation projects to be announced, VERITAS, has yet to break
ground, first light in 2005 is still feasible. The space
telescopes, reported elsewhere in these proceedings, seem to be
well on schedule. The scheduled completion date for MACE, the
Indian MAGIC look-alike, is 2003.

Although AMS \cite{AMS99} and AGILE \cite{AGILE99} will
partially fill the HE gap left by the demise of EGRET, they will
not provide a significant boost in sensitivity. In the next few
years the most exciting new results may well come from the solar
telescopes as they explore a new energy domain. Prior to the launch
of GLAST \cite{GLAST99} there will be a spate of new results
forthcoming from the new generation of IACT arrays.

{\bf Acknowledgements:}
Research in very high energy gamma-ray astronomy at the Smithsonian
Astrophysical Observatory is supported by the U.S. Department of
Energy. I am grateful to Tony Hall, Deirdre Horan, Jim Gaidos and Rod 
Lessard  for helpful comments
on the manuscript.


\begin{thebibliography}{}

\bibitem{G&J53} Galbraith, W., Jelley, J.V., Nature, {\bf 171}, 349
(1953)

\bibitem{G&J57} Jelley, J.V., Galbraith, W,., J. Atmos. Phys. {\bf
6}, 
304 (1955)

\bibitem{chudakov65} Chudakov, A.E et al., Trans. Consult. Bureau,
P.N.Lebedev Inst., {\bf 26}, 99 (1965)

\bibitem{J&P63} Jelley, J.V., Porter, N.A., M.N.R.A.S. {\bf 4}, 275
(1963)

\bibitem{morrison58} Morrison, P., Nuovo Cimento, {\bf 7}, 858
(1958)

\bibitem{cocconi60} Cocconi, G., 6th ICRC, (Moscow), {\bf 2}, 309
(1959)

\bibitem{weekes99}
Weekes, T.C. "GeV-TeV Gamma Ray Astrophysics Workshop", Snowbird,
UT. August, 1999. AIP Proc. Conf. {\bf 515}, 3

\bibitem{buckley99} Buckley, J.H. 26th ICRC, (Salt Lake City), AIP
Conference
Proceedings{\bf 516}, Eds. B.Dingus, D.Kieda, M.Salamon 195 (1999)

\bibitem{gaidos96} Gaidos, J.A. et al., Nature, {\bf 383}, 319
(1996)

\bibitem{biller99} Biller, S. et al., Phys. Rev. Lettr. {\bf 83},
2108
(1999)

\bibitem{kranich99} Kranich, D. et al., 26th ICRC, (Salt Lake
City),
 {\bf 3}, 358 (1999) 

\bibitem{takahashi00} Takahashi, T. et al., Ap. J. (in press)
(2000)

\bibitem{cataneseweekes99} Catanese, M., Weekes, T.C., P.A.S.P.
{\bf
111}, 1193 (1999)

\bibitem{maraschi99} Maraschi, L. et al., Astropart. Phys. {\bf
11}, 189
(1999)

\bibitem{aharonian99} Aharonian, F.A. et al., A. \& A., {\bf 349},
11
(1999)

\bibitem{WHIPPLE} Cawley, M.F. et al., Exp. Ast. {\bf 1}, 185
(1990)

\bibitem{CRIMEA} Vladimirsky, B.M. et al., Proc. Workshop on VHE
Gamma
Ray Astronomy, Crimea (April, 1989), 21 (1989)

\bibitem{SHALON} Nikolsky S.I, Sinitsyna V.G., Proc. Workshop on
VHE Gamma Ray Astronony, Crimea (April, 1989), 11 (1989)

\bibitem{CANGAROO} Hara, T. et al., Nucl.Inst. Meth. A. {\bf
332}, 300 (1993)

\bibitem{HEGRA}  Daum, A. et al., Astropart. Phys. {\bf 8}, 1
(1997)

\bibitem{CAT} Goret, P. et al., 25th ICRC, (Durban), {\bf 3}, 173
(1997)

\bibitem{TACTIC} Bhat, C.L. et al., Proc. Workshop on VHE Gamma Ray
Astronomy, (Kruger Park), (August, 1997) 196 (1997)

\bibitem{DURHAM99} Chadwick, P.M. et al., Proc. Workshop on VHE
Gamma Ray
Astronomy, (Snowbird), (August, 1999), 223 (1999)

\bibitem{7TEL99}
T.Yamamoto et al., 26th ICRC, (Salt Lake City), {\bf 5}, 275 (1999)

\bibitem{POTCH} 
DeJager, H.I. et al., South African of J. Phys., {\bf9} 107 (1986)

\bibitem{PACHMARI99}
Bhat, P.N. et al., 26th ICRC, (Salt Lake City), {\bf 5}, 191 (1999)

\bibitem{BEIJING} 
Yinlin, J. et al., 21st ICRC, (Adelaide), {\bf 4}, 220 (1990)

\bibitem{STACEE99} 
Ong, R.A. "GeV-TeV Gamma Ray Astrophysics Workshop", Snowbird, UT.
August, 1999. AIP Proc. Conf. {\bf 515}, 401.

\bibitem{CELESTE99}
de Naurois, M., 26th ICRC, (Salt Lake City),  {\bf 5}, 211 (1999)

\bibitem{SOLAR-299}
Zweerink, J.A. et al., 26th ICRC, (Salt Lake City), {bf 5}, 223
(1999)

\bibitem{GRAAL99}
Arqueros, F. et al., 26th ICRC, (Salt Lake City), {\bf 5}, 215
(1999).

\bibitem{MILAGRO} Yodh, G.B., Space. Sci. Rev. {\bf 75}, 199 (1996)


\bibitem{TIBET99} Amenomori, M. et al. 
"GeV-TeV Gamma Ray Astrophysics
Workshop", Snowbird, UT. August, 1999. AIP Proc. Conf. {\bf 515},
459 (1999) 


\bibitem{mori99} Mori, M. et al. "GeV-TeV Gamma Ray Astrophysics
Workshop", Snowbird, UT. August, 1999. AIP Proc. Conf. {\bf 515},
485

\bibitem{HESS99}
Kohnle, A. et al., 26th ICRC, {Salt Lake City}, {\bf 5}, 239 (1999)

\bibitem{VERITAS99} 
Bradbury, S.M. et al., 26th ICRC, {\bf 5}, 280 (1999)

\bibitem{MAGIC99}
Martinez, M., 26th ICRC, (Salt Lake City),  {\bf 5}, 219 (1999).

\bibitem{hofmann99} Hofmann, W. 
"GeV-TeV Gamma Ray Astrophysics
Workshop", Snowbird, UT. August, 1999. AIP Proc. Conf. {\bf 515}, 
492 (1999)



\bibitem{vassiliev99} 
Vassiliev, V.V. et al., 26th ICRC,  (Salt Lake City), {\bf 5}, 299
(1999)

\bibitem{AMS99}
Mereghetti, S. et al. "GeV-TeV Gamma Ray Astrophysics Workshop",
Snowbird, UT. August, 1999. AIP Proc. Conf. {\bf 515}, 467 (1999) 


\bibitem{AGILE99}
Battiston, R. "GeV-TeV Gamma Ray Astrophysics Workshop", Snowbird,
UT. August, 1999. AIP Proc. Conf. {\bf 515},  474 (1999)


\bibitem{GLAST99}
Kniffen, D.A., Bertsch, D.L. and Gehrels, N. "GeV-TeV Gamma Ray
Astrophysics Workshop", Snowbird, UT. AIP Proc. Conf.
{\bf 515},  492 (1999)

\end{thebibliography}
\end{document}